\documentclass[letterpaper,11pt]{article}
\usepackage{tensor}
\usepackage{calligra}
\usepackage{physics}
\usepackage[]{hyperref}
\usepackage{amsmath}
\usepackage{amsfonts}
\usepackage{amssymb}
\usepackage{color}
\usepackage{mathrsfs}
\usepackage{latexsym}
\usepackage{physics}
\usepackage{cite}
\usepackage{graphicx}
\usepackage{float}
\usepackage{leftidx}
\usepackage{xcolor}
\graphicspath{ {images/} }
\title{Bending of light in the Universe filled with quintessential dark energy}
\author{R. Saadati$^1$, F. Shojai$^{1,2}$,\\$^1$Department of Physics, University of Tehran,\\Tehran, Iran.\\$^2$Foundations of Physics Group, School of Physics,\\Institute for Research in Fundamental Sciences (IPM),\\Tehran, Iran.\\}
\date{}
\begin{document}
\maketitle
\begin{abstract}
As a local effect of dynamical dark energy, bending of light in the presence of a spherically symmetric and static black hole surrounded by quintessence has been studied. Having in mind recent observational data, we have treated the problem as a deviation from Kottler space-time. This deviation is measured by a perturbation parameter $\varepsilon$ included in the equation of state parameter of quintessence as $\omega_q=-1+\frac{1}{3}\varepsilon$.  Here, the deflection angle is calculated and then the result  is compared with \cite{Arakida:2011th} in the limit $\varepsilon\rightarrow 0$ where the quintessence behaves like the cosmological constant. It is shown that unlike the cosmological constant, the effect of quintessence on the photon energy equation can not be absorbed into the definition of impact parameter. Moreover in this paper, we  generalize the Kiselev black hole to the case that there is a modified Chaplygin gas as the dark energy component of the universe and show that the resulted metric can be reduced to the Kiselev metric by adjusting some arbitrary parameters. 
\end{abstract}
\section{Introduction}
The universe undergoes an accelerated expansion which is caused by a peculiar relativistic agent, called dark energy whose presence gives rise to repulsive gravity. The most famous and simplest candidate of dark energy is the cosmological constant which has constant energy density and pressure. There are also some other candidates with varying energy density. One of which is called quintessence with an equation of state parameter in the range of $-1<\omega_q<-1/3$.  For a comprehensive review on dark energy dynamics see\cite{Copeland:2006wr}. The observational supports for the presence of dark energy are many, see for example \cite{R} for type Ia supernovae observations and also \cite{amen} for a summary of the current status of dark energy research. Problems of cosmological constant, such as the conflict between the obtained value of the vacuum energy density predicted by quantum field theory and its cosmological observed value, lead physicists to consider the dynamical dark energy models. Recent observational data, see for example Planck 2018 results \cite{Agha} and DES Collaboration results \cite{abb}, indicate that the value of dark energy state parameter, $\omega$ lies in a narrow strip around $-1$ \cite{alam}.

Besides the cosmological effects of dark energy, it has some local consequences. The effect of cosmological constant in bending of light is studied in\cite{Arakida:2011th} and \cite{Rindler:2007zz}, although in some earlier works, it had been argued that the cosmological constant has nothing to do with the bending of light, see for example \cite{Kerr:2003bp}.

On the other hand in 2003 Kiselev introduced an exact solution of Einstein field equations for a static black hole surrounded by quintessence \cite{Kiselev:2002dx} which enabled the investigation of the local consequences of quintessence. Null geodesics in Kiselev metric has been examined in detail for special case $\omega_q=-2/3$ in\cite{Fernando:2012ue}. For this case, the effect of quintessence on photon trajectory is examined in \cite{Malakolkalami:2015cza}. Also, the study of null geodesics around a charged black hole immersed in quintessence is done in \cite{Fernando:2014rsa}. Using Janis-Newman algoritm, the Kiselve metric is generalized to kerr \cite{Toshmatov} and to Kerr-Newmann-Ads \cite{new} black holes. Also, the authors of \cite{Rizwan:2018lht} study the critical values of quintessential and spin parameters, to distinguish a rotating
Kiselev black hole from a naked singularity. Other works in this topic can be founded in papers \cite{Uniyal:2014paa,Younas:2015sva}. 

In this paper we are going to treat the bending of light in Kiselve space-time where the equation of state parameter of quintessence is $\omega_q=-1+\frac{1}{3}\varepsilon$ in which $\vert\varepsilon\vert\ll1$. The coefficient $\frac{1}{3}$ comes for convenience. This parameter can be considered as a perturbation parameter. In the limit of $\varepsilon\rightarrow0$, the results of Kottler (Schwarzschild-de Sitter) space-time  will be obtained \cite{Arakida:2011th}. Kiselev metric will be our starting point and after having a review about it in the next section, we take a glance on the results of bending of light, obtained in the Kottler space-time, in section \ref{bend}. Then the photon energy equation and the deflection angle will be calculated in the Kiselev space-time perturbatively in section \ref{sec: Our Contribution}. In section \ref{Sec: Concluding Remarks} we will generalize the Kiselev metric for a black hole immersed in the Chaplygin gas and finally we will give some conclusions in section \ref{Sec: last}.
\section{Kiselev space-time}\label{Kiselev Space-time}
In a static spherically symmetric space-time, the most general form of the line element can be written as:
\begin{equation}\label{metric: caconical form}
ds^2=e^{\nu(r)}dt^2-e^{\lambda(r)}dr^2-r^2(d\theta^2+\sin^2\theta d\phi^2)
\end{equation}
Using  Einstein equations, one can obtain the energy-momentum tensor components as follows ($4\pi G=1$):
\begin{equation}\label{T_t^t}
T\indices{_t^t}=-\frac{1}{2}e^{-\lambda}\left(\frac{1}{r^2}-\frac{\lambda'}{r}\right)+\frac{1}{2r^2}
\end{equation}
\begin{equation}\label{T_r^r}
T\indices{_r^r}=-\frac{1}{2}e^{-\lambda}\left(\frac{1}{r^2}+\frac{\nu'}{r}\right)+\frac{1}{2r^2}
\end{equation}
and
\begin{equation}
T\indices{_\theta^\theta}=T\indices{_\phi^\phi}=-\frac{1}{4}e^{-\lambda}\left(\nu''+\frac{\nu'^2}{2}+\frac{\nu'-\lambda'}{r}-\frac{\nu'\lambda'}{2}\right)
\end{equation}
where primes denote differentiation with respect to the radial coordinate. In the case $\nu'+\lambda'=0$, we see that the first two equations lead to 
\begin{equation}\label{linear condition}
T\indices{_t^t}=T\indices{_r^r}
\end{equation}
Substituting $\lambda=-\ln\left(1+f\right)$, it can be easily seen that Einstein equations reduce to some linear equations for the unknown function $f(r)$:  
\begin{equation}\label{linear1}
T\indices{_t^t}=T\indices{_r^r}=-\frac{1}{2r^2}\left(f+rf'\right)
\end{equation}
\begin{equation}\label{linear2}
T\indices{_\theta^\theta}=T\indices{_\phi^\phi}=-\frac{1}{4r}\left(2f'+rf''\right)
\end{equation}
and thus the superposition principle would be satisfied \cite{pad}.\\
Following \cite{Kiselev:2002dx},  for a static spherically symmetric space-time, the components of the energy-momentum tensor in cartesian coordinates are of the form:
\begin{equation}
T\indices{_\mu^\nu}=\left(\begin{array}{cc}
A(r) & 0 \\
\\
0    & C(r)r_i r^j+B(r)\delta\indices{_i^j}\\
\end{array}\right)
\end{equation}
Averaging over angles leads to
\begin{equation}\label{T averaged}
\langle T\indices{_\mu^\nu}\rangle=\text{diag}\bigg(A(r), \frac{1}{3}r^2C(r)+B(r),\frac{1}{3}r^2C(r)+B(r),\frac{1}{3}r^2C(r)+B(r)\bigg )
\end{equation}
Comparing this with the energy-momentum tensor of a perfect fluid
\begin{equation}
T\indices{_\mu^\nu}=\text{diag}\big ( \rho(r), -p(r),-p(r),-p(r) \big )
\end{equation}
we see that:
	\begin{equation} \label{ABCa}
	A(r)=\rho(r)
	\end{equation}
	\begin{equation}\label{ABCb}
	\frac{1}{3}r^2C(r)+B(r)=-p(r)
	\end{equation}
	in which $\rho(r)$ and $p(r)$ are the energy density and pressure respectively. Performing a coordinate transformation, it is straightforward to show that the  energy-momentum tensor in spherical coordinates is given by
\begin{equation}\label{r}
T\indices{_{\mu'}^{\nu'}}=\text{diag}\big( \rho(r), C(r)r^2+B(r), B(r), B(r)\big)
\end{equation}
Applying the condition \eqref{linear condition} yields:
\begin{equation}\label{akhr}
C(r)r^2+B(r)=\rho(r)
\end{equation}
Equations (\ref{ABCa}), (\ref{ABCb}) and (\ref{akhr}) specify the energy-momentum tensor in spherical coordinates as
\begin{equation}\label{tt}
T\indices{_{\mu'}^{\nu'}}=\text{diag}\bigg ( \rho(r), \rho(r), -\frac{1}{2}(\rho(r)+3 p(r)), -\frac{1}{2}(\rho(r)+3 p(r))\bigg )
\end{equation}
It is worth noting that the above energy-momentum tensor is not of the form of a perfect fluid except for the cosmological constant where  $p=-\rho$. It presents an anisotropic fluid with the following radial and transverse pressures 
\begin{equation}\label{pha}
p_r=-\rho(r)  \hspace{1cm} p_t=\frac{1}{2}(\rho(r)+3p(r))
\end{equation}
Using (\ref{tt}), the Einstein's field equations \eqref{linear1} and \eqref{linear2} read
\begin{equation}\label{e1}
-\frac{1}{2r^2}\left(f+rf'\right)=\rho(r)
\end{equation}
\begin{equation}\label{e2}
-\frac{1}{4r}\left(2f'+rf''\right)= -\frac{1}{2}\left(\rho(r)+3p(r)\right)
\end{equation}
Assuming a linear equation of state, $p(r)=\omega\rho(r)$, from above two equations, one quickly find that 
\begin{equation}\label{eq: f_q}
f=-\frac{r_g}{r}-\left(\frac{r_q}{r}\right)^{3\omega+1}
\end{equation}
\begin{equation}
p_r(r)=-\frac{3\omega}{2r^2}\left(\frac{r_q}{r}\right)^{3\omega+1} \hspace{1cm} p_t(r)=-\frac{3\omega(1+3\omega)}{4r^2}\left(\frac{r_q}{r}\right)^{3\omega+1}
\end{equation}
where $r_g$ is the Schwarzschild radius and $r_q$ is another integration constant with length dimension.  To find the order of magnitude of $r_q$, setting\\ $\omega_q =-1$, the  temporal component of the metric takes the form
\begin{equation}\label{kot}
g_{tt}=1-\frac{r_g}{r}-\left(\frac{r}{r_q}\right)^2
\end{equation}
which is Kottler or Schwarzschild-de Sitter space-time with cosmological constant
\begin{equation}\label{r_q}
\Lambda=\frac{3}{r_q^2}
\end{equation}
This means that $r_q$ is of the order of the de Sitter radius. Using the superposition principle, the solution for different values of state parameter can be superposed and the corresponding metric component is:
\begin{equation}\label{eq: f_q Sum}
g_{tt}=1-\frac{r_g}{r}-\sum_{n}\left(\frac{r_n}{r}\right)^{3\omega_n+1}.
\end{equation}
which is known as Kiselev space-time \cite{Kiselev:2002dx}. 
\section{Bending of light in Kottler space-time}\label{bend}
In the absence of cosmological constant and other types of quintessence, the Kiselev metric reduces to the Schwarzschild one. Solving the null geodesic equation in the equatorial plane in Schwarzschild space-time, the photon trajectory and the deflection angle up to the second order in $r_g$  are\cite{Epstein:1980dw}:
\begin{equation}\label{traj b}
\frac{1}{r}=\frac{1}{b}\sin\varphi+\frac{r_g}{4b^2}(3+\cos2\varphi)+\frac{r_g^2}{64b^3}\left(37\sin\varphi+30(\pi-2\varphi)\cos\varphi-3\sin3\varphi\right)
\end{equation}
\begin{equation}\label{eq:deflection angel b}
\delta_S=2\frac{r_g}{b}+\frac{15\pi}{16}\left(\frac{r_g}{b}\right)^2
\end{equation}
where $b$ is the impact parameter which determines the shortest spatial distance to the origin. In deriving the above relations, the integration constants are chosen such that the radial distance is minimized where $\varphi=\pi/2$ at any order of expansion. 

The effect of cosmological constant $\Lambda$ on the photon trajectory and the bending of light is also examined in \cite{Arakida:2011th}. In Kottler space-time (\ref{kot}), the geodesic equation in equatorial space-time can be written as
\begin{equation}\label{traje}
\left (\frac{d\bar{u}}{d\varphi}\right)^2=1-\bar{u}^2+\bar{r}_g\bar{u}^3+\frac{\bar{\Lambda}}{3}
\end{equation}
where for convenience, we have used the dimensionless variables $\bar{r}=r/b$, $\bar{u}=1/\bar{r}$, $\bar{\Lambda}=b^2\Lambda$ and $ \bar{r}_g=r_g/b $. Comparing with the Schwarzschild case, the effect of cosmological constant is equivalent to defining a new dimensionless impact parameter $\bar{B}$ which is
\begin{equation}\label{eq: Impact parameter B}
\frac{1}{\bar{B}^2}\equiv 1+\frac{\bar{\Lambda}}{3}
\end{equation}
This means that the first order differential equation of the light path changes when the cosmological constant is included \cite{Arakida:2011th}.
Thus to include the cosmological constant, it is sufficient to replace $b$ and $r_g$ with $\bar{B}$ and $\bar{r}_g$  in relation (\ref{traj b}) to obtain the following photon trajectory
\begin{equation}\label{traj}
\frac{1}{\bar{r}}=\frac{1}{\bar{B}}\sin\varphi+\frac{\bar{r}_g}{4\bar{B}^2}(3+\cos2\varphi)+\frac{\bar{r}_g^2}{64\bar{B}^3}\left(37\sin\varphi+30(\pi-2\varphi)\cos\varphi-3\sin3\varphi\right)
\end{equation}
but it is not correct to do so for the deflection angle (\ref{eq:deflection angel b}). To clarify this point more, let us use the method of \cite{Rindler:2007zz} in which one should calculate the inner product of two coordinate directions in curved space-time. The first 
is the direction of photon propagation in the two dimensional space $(r,\phi)$ and the other is the coordinate line corresponding to the constant azimuth angle. By a simple calculation one can show that the angle between these two directions is \cite{Rindler:2007zz}
\begin{equation}\label{PSI}
\tan \psi=\frac{\sqrt{g_{tt}(\bar{r})}\bar{r}}{|d\bar{r}/d\varphi|}
\end{equation}
and the one-sided deflection angle is defined as $\delta_\Lambda=\psi-\varphi$ for arbitrary constant $\varphi$. This angle is very small thus $\delta_\Lambda\simeq\tan\delta_\Lambda=\tan(\psi-\varphi)$. Substituting (\ref{eq: Impact parameter B}) and (\ref{traj}) into (\ref{PSI}) and expanding the result up to the first order in $\bar{\Lambda}$ and the second order in $\bar{r}_g$, gives 
\[
\tan{\psi}=
\left (\tan{\varphi}-\frac{\bar{\Lambda}}{3 \sin2\varphi}\right)+ \left(\frac{1}{\cos{\varphi}}+\frac{\bar{\Lambda} \cos{\varphi}}{6 \sin^2{\varphi}}\right)\bar{r}_g+
\]
\[
\left(\frac{15(\pi-2\varphi+\sin{2\varphi})}{32\cos^2{\varphi}}-\frac{\bar{\Lambda}}{96 \sin\varphi \sin^2 2\varphi}\times\right .
\]
\begin{equation}\label{ppssii}
\Big(33\cos\varphi+31\cos 3\varphi -30(\pi-2\varphi)\sin\varphi\Big)\bigg)\bar{r}_g^2
\end{equation} 
According to (\ref{kot}) and (\ref{r_q}), if the radial coordinate exceeds from the dimensionless 
de Sitter horizon, $\sqrt{3/\bar{\Lambda}}$, the cosmological constant considerably affects on the bending of light path. 
Thus, at small enough $\varphi$ angle, in which $r$ can be compared with the horizon value, the cosmological constant can have a significant effect on the $\psi$ angle and thus on the deflection angle of photons. One can easily show that replacing $b$ and $r_g$ with $\bar{B}$ and $\bar{r}_g$ in relation (\ref{eq:deflection angel b})(see relation (13) in \cite{Arakida:2011th}) and then expanding it, the resulted expression for deflection angle is different from what can be obtaind by (\ref{ppssii}).

To compare the Kottler bending light with the Schwarzschild one, one can compute the relative difference in bending angle, $\Delta_\Lambda=(\delta_\Lambda-\delta_S)/\delta_S$ in two cases up to any arbitrary order of expansion. Up to the first order in $\bar r_g$ and $\bar \Lambda$, one gets
\begin{equation}
\bar r_g\Delta_\Lambda=-\frac{1}{6\sin\varphi}\bar \Lambda+\left(\frac{1}{6\sin^2\varphi}+\frac{5}{32}\left(1+\frac{\pi-2\varphi}{\sin2\varphi}\right)\right)\bar r_g\bar \Lambda
\end{equation}
\section{Bending of light in Kiselev space-time}\label{sec: Our Contribution}
Now let us to consider the deflection of light in Kiselev space-time given by (\ref{eq: f_q}). Using the null geodesic equations in the equatorial plane,
one can write the photon energy equation as
\begin{equation}\label{eq: Geodeisc1}
\left(\dv{\bar{u}}{\varphi}\right)^2=1-\bar{u}^2+\bar{r}_g \bar{u}^3+\bar{r_q}^{\varepsilon-2} \bar{u}^{\varepsilon}
\end{equation} 
where $\bar{r_q}=\sqrt{3/\bar{\Lambda}}$. Equation \eqref{eq: Geodeisc1} differs from the photon geodesic equation in the Schwarzschild space-time in the last term on its right hand side which is the quintessential correction. Here we consider the Kiselev metric as a nearly Kottler metric if the equation of state parameter of quintessence is near $-1$. Hence, we write $\omega_q=-1+\frac{1}{3}\varepsilon$ in which the small parameter $\varepsilon$  measures the deviation of the metric from the Kottler metric. The time-time component of metric, (\ref{eq: f_q Sum}), up to the first order of $\varepsilon$ would be
\begin{equation}\label{ff}
g_{tt}=1-\frac{\bar{\Lambda}}{3\bar{u}^2}-\bar{r}_g\bar{u}-\varepsilon\frac{\bar{\Lambda}}{3\bar{u}^2}\ln {\sqrt{\frac{3}{\bar{\Lambda}}}\bar{u}}
\end{equation}
This means that if the  quintessential dark energy is considered to be a small deviation from the cosmological constant, it affects as a perturbation on the Kottler space-time. This is also true for the null geodesic equation considered here. Expanding the last term of \eqref{eq: Geodeisc1} up to the first order in $\varepsilon$ leads
\begin{equation}\label{eq: Geodeisc}
\left(\dv{\bar{u}}{\varphi}\right)^2=1-\bar{u}^2+\bar{r}_g \bar{u}^3+\frac{\bar{\Lambda}}{3}+\frac{\varepsilon\bar \Lambda}{6}\ln{\frac{3}{\bar{\Lambda}}\bar{u}^2}
\end{equation}
where the last term on the right hand side of  (\ref{eq: Geodeisc}) is a perturbation to the Kottler null geodesic equation (\ref{traje}). In the following, in addition to $\bar r_g$, we use another expansion parameter $\varepsilon$  . Then we expand $\bar u(\varphi)$ in terms of these small parameters as follows:
\begin{equation}\label{eq: u_0(phi)}
\bar{u}=\left[(\bar{u_0}+\varepsilon\bar{w_0}+...)+\bar{r}_g(\bar{u_1}+\varepsilon \bar{w_1}+...)+\bar{r}_g^2(\bar{u_2}+\varepsilon\bar{w_2}+...)+...\right]
\end{equation}
where, on the right-hand side, any terms in $\bar r_g$ Taylor expansion, is also expanded in terms of $\varepsilon$. 
 
Inserting \eqref{eq: u_0(phi)} into \eqref{eq: Geodeisc}, we find that the functions $\bar u_0$, $\bar u_1$, $\bar u_2$, $\bar w_0$, $\bar w_1$ and $\bar w_2$ satisfy the following equations:
\begin{equation}
\left(\frac{d\bar{u_0}}{d\varphi}\right)^2=1-\bar{u_0}^2+\frac{\bar{\Lambda}}{3}
\end{equation}
\begin{equation}\label{u1}
\left(\frac{d\bar{u_0}}{d\varphi}\right)\left(\frac{d\bar{u_1}}{d\varphi}\right)+\bar{u_0}\bar{u_1}=\frac{1}{2}u_0^3
\end{equation}
\begin{equation}\label{u2}
\frac{1}{2}\left(\frac{d\bar{u_1}}{d\varphi}\right)^2+\left(\frac{d\bar{u_0}}{d\varphi}\right)\left(\frac{d\bar{u_2}}{d\varphi}\right)=+3\bar{u_0}^2\bar{u_1}-\bar{u_0}\bar{u_2}-\frac{1}{2}\bar{u_1}^2
\end{equation}
\begin{equation}\label{w0}
\left(\frac{d\bar{u_0}}{d\varphi}\right)\left(\frac{d\bar{w_0}}{d\varphi}\right)+\bar{u_0}\bar{w_0}=\frac{\bar{\Lambda}}{12}\ln({\frac{3}{\bar{\Lambda}}\bar{u_0}^2})
\end{equation}
\begin{equation}\label{w1}
\left(\frac{d\bar{u_0}}{d\varphi}\right)\left(\frac{d\bar{w_1}}{d\varphi}\right)+\bar{u_0}\bar{w_1}=3\bar{u_0}^2\bar{w_0}
-\left(\frac{d\bar{u_1}}{d\varphi}\right)\left(\frac{d\bar{w_0}}{d\varphi}\right)-\bar{u_1}\bar{w_0}+\frac{\bar{\Lambda}}{6}\frac{\bar{u_1}}{\bar{u_0}}
\end{equation}
\[
\left(\frac{d\bar{u_0}}{d\varphi}\right)\left(\frac{d\bar{w_2}}{d\varphi}\right)+\bar{u_0}\bar{w_2}=3\bar{u_0}\bar{u_1}\bar{w_0}+\frac{3}{2}\bar{u_0}^2\bar{w_1}-\left(\frac{d\bar{u_1}}{d\varphi}\right)\left(\frac{d\bar{w_1}}{d\varphi}\right)\]
\begin{equation}\label{w2}
-
\left(\frac{d\bar{u_2}}{d\varphi}\right)\left(\frac{d\bar{w_0}}{d\varphi}\right)-
\bar{u_1}\bar{w_1}-\bar{u_2}\bar{w_0}+\frac{\bar{\Lambda}}{6}(\frac{\bar{u_2}}{\bar{u_0}}-\frac{\bar{u_1}^2}{2\bar{u_0}^2})
\end{equation}
The first three equations are exactly those are obtained from the Kottler space-time in \cite{{Arakida:2011th}} and their solution yields (\ref{traj}). Substituting (\ref{eq: u_0(phi)}) into (\ref{w0}-\ref{w2}) and assuming the minimum of $r$ occurs  at $\varphi=\pi/2$ at any order of expansion, we find the photon trajectory as following \footnote{It should be noted that the last parentheses in (\ref{W3}) comes from simplifying $i(\pi^2/6-\varphi^2)+2\varphi\ln(1-e^{2i\varphi})-i Li_2\left(e^{2i\phi}\right)$. A simple calculation shows that this expression is real. To see this, it is enough to use: $\Im \ln(1-e^{2i\varphi})=\varphi-\pi/2$ and $\Re Li_2\left(e^{2i\phi}\right)=\pi^2/6-\pi\varphi+\varphi^2$.}:
\[
\frac{1}{\bar{r}}=\left[\frac{1}{\bar{B}}\sin\varphi+\varepsilon\frac{\bar{B}\bar{\Lambda}}{12}\Bigg((\pi-2\varphi)\cos\varphi+\sin\varphi\ln(\frac{3}{\bar{\Lambda}\bar{B}^2}\sin^2\varphi)\Bigg)\right]
\]
\[
+\bar{r}_g\left[\frac{1}{4\bar{B}^2}(3+\cos2\varphi)+\varepsilon\frac{\bar{\Lambda}}{12}\Bigg(1+\left(1+\cos^2\varphi\right)\ln(\frac{3}{\bar{\Lambda}\bar{B}^2}\sin^2\varphi)\right.
\]
\[
-\left.\cos\varphi\ln(\tan^2{\frac{\varphi}{2}})-\frac{\pi-2\varphi}{2}\sin 2\varphi\Bigg) \right]
\]
\[
+\bar{r}_g^2\left[\frac{1}{64\bar{B}^3}\Bigg(37\sin\varphi+30(\pi-2\varphi)\cos\varphi-3\sin3\varphi\Bigg)\right .
\]
\[
+\varepsilon\frac{\bar{\Lambda}}{1536\bar{B}\sin\varphi}\left\{2\cos 2\varphi \Bigg(-32+15(\pi-2\varphi)^2-60\ln\left(\frac{3}{\bar{B}^2\bar{\Lambda}}\sin^2\varphi \right)\Bigg)
\right .
\]
\[
 + 3\Bigg(64-10(\pi-2\varphi)^2+ 37\ln\left(\frac{3}{\bar{B}^2\bar{\Lambda}}\sin^2\varphi \right)\Bigg)+128\cos\varphi\ln\left(\tan^2\frac{\varphi}{2}\right)\sin^2\varphi
\]
\[
+10\Bigg[-34\varphi+ \pi(17-12\ln 2)+3\pi\ln \left(\left(\frac{3}{\bar{B}^2\bar{\Lambda}}\right)^3\sin^2\varphi \right)\Bigg]\sin 2\varphi
\]
\[
+9\ln\left(\frac{3}{\bar{B}^2\bar{\Lambda}}\sin^2\varphi \right)\left(\cos 4\varphi -20\varphi\sin2\varphi\right)+(18\varphi-9\pi)\sin 4\varphi
\]
\begin{equation}\label{W3}
\left . +120\sin 2\varphi\Bigg ( 2\varphi\Re\ln(1-e^{2i\varphi})+\Im Li_2\left(e^{2i\varphi}\right) \Bigg )\right \} \Bigg]
\end{equation}
Substituting (\ref{ff}) and (\ref{W3}) into (\ref{PSI}) gives the $\psi$ angle. For a more accurate comparison with (\ref{ppssii}), we expand the result up to the first order in $\bar\Lambda$ 
\[
\tan\psi=\tan\varphi+\frac{\bar r_g}{\cos{\varphi}}-\frac{\bar\Lambda}{3\sin2\varphi}+\frac{\cos{\varphi}}{6 \sin^2{\varphi}}\bar{\Lambda} \bar{r}_g-
\]
\[\frac{1}{6}\left(\frac{2\varphi-\pi}{2\cos^2\varphi}-\frac{1}{\sin^2\varphi}\ln\frac{\bar\Lambda}{3\sin^2\varphi}\right)\varepsilon\bar\Lambda 
-\frac{1}{12}\left[\frac{1}{\cos^2\varphi}\left(2\ln\tan\frac{\varphi}{2}+(2\varphi-\pi)\sin\varphi\right)+\right .
\]
\[
\left.\frac{\cos\varphi}{\sin^2 2\varphi}\left(8+\cos3\varphi+\frac{3(\cos\varphi+1)}{\cos\varphi}\ln{\frac{\bar\Lambda}{3\sin^2\varphi}}\right)\right]\bar r_g\varepsilon\bar \Lambda-\frac{15(2\varphi-\pi-\sin2\varphi)}{32\cos^2\varphi}\bar r_g^2-
\]
\[
\frac{1}{96\sin\varphi\sin2\varphi}\bigg(33\cos\varphi+31\cos3\varphi+30(2\varphi-\pi)\sin\varphi\bigg)\bar r_g^2\bar\Lambda-
\]
\[
\frac{1}{1536\sin^32\varphi}\left\{-478+340(2\varphi-\pi)-360 (2\varphi-\pi)^2+120\pi\ln 2+1113\ln\frac{\bar\Lambda}{3\sin^2\varphi}+\right .
\]
\[
64\sin2\varphi\sin^2\varphi\left(16\sin\varphi\ln\tan\frac{\varphi}{2}-15\Im Li_2(e^{2i\varphi})\right)+
\]
\[
2\cos2\varphi\left(-772+240(2\varphi-\pi)^2+499\ln\frac{\bar\Lambda}{3\sin^2\varphi}\right)-
\]
\[
2\cos4\varphi\left (529+170(2\varphi-\pi)+60(2\varphi-\pi)^2+120\pi\ln2+853\ln\frac{\bar\Lambda}{3\sin^2\varphi} \right )+
\]
\[
2\cos6\varphi\left (4+243\ln\frac{\bar\Lambda}{3\sin^2\varphi} \right )+81\cos8\varphi \ln\frac{\bar\Lambda}{3\sin^2\varphi}-
\]
\[
4\sin2\varphi\left (170-187\pi+1589\varphi+240\varphi\ln2+60(2\varphi-\pi)\ln\frac{\bar\Lambda}{3}+1620\varphi\ln\frac{\bar\Lambda}{3\sin^2\varphi}  \right)-
\]
\[
10\sin4\varphi\left (-34-48\varphi\ln2+19(2\varphi-\pi)+12\varphi\ln\frac{\bar\Lambda}{3}-336\varphi\ln\frac{\bar\Lambda}{3\sin^2\varphi} \right)+
\]
\begin{equation}
324(7\varphi-\pi)\sin6\varphi-81(2\varphi-\pi)\sin8\varphi\bigg\}\bar r^2_g\varepsilon\bar \Lambda
\end{equation}
This shows how deviation from the cosmological constant affects on the deflection angle. For completeness analysis, one can find  $\Delta_{\Lambda\varepsilon}= (\delta_{\Lambda\varepsilon}-\delta_\Lambda)/\delta_\Lambda$ up to the first order in $r_g$ , $\Lambda$  and $\varepsilon$ as follows
\[
r_g\Delta_{\Lambda \varepsilon}=\frac{1}{12}\left(\frac{\pi-2\varphi}{\cos\varphi}+\frac{1}{\sin\varphi}\ln\frac{\bar\Lambda}{3\sin^2 \varphi}\right)\varepsilon\bar\Lambda-
\]
\[
\frac{1}{384\sin^2 2\varphi}\left[128+30(\pi-2\varphi)^2+30(\pi-2\varphi)\sin2\varphi\left(\sin^2\varphi+\ln{\frac{\bar \Lambda}{3\sin^2\varphi}}\right)+\right.
\]
\[
128\sin\varphi\sin2\varphi\ln\tan\frac{\varphi}{2}+79\ln\frac{\bar \Lambda}{3\sin^2\varphi}-2\cos^2\varphi\left(15(\pi-2\varphi)^2-\right.
\]
\[
\left.\left.64-32\ln\frac{\bar \Lambda}{3\sin^2\varphi}\right)-15\cos4\varphi\ln\frac{\bar \Lambda}{3\sin^2\varphi}\right]\bar r_g\bar \Lambda\varepsilon
\]

\section{Black holes in the presence of Chaplygin gas}\label{Sec: Concluding Remarks}
Although in this paper we have assumed that the dark energy equation of state is very close to the cosmological constant, it is known that there are several candidate for dark energy in addition to the cosmological constant and the quintessence such as the chaplygin gas, phantom, k-essence and so on \cite{Copeland:2006wr}. Thus, it is of some interest to find the black hole solutions of general relativity surrounded by an arbitrary kind of dark energy. Here, for an example, we are going to do this by accepting the Chaplygin gas \cite{Kamenshchik:2001cp,Bento:2002ps} as the dark energy component of the universe.

For a modified Chaplygin gas, the pressure $p$ and the energy density $\rho$ are related through the following equation of state:
\begin{equation}\label{Chaplygin State Equation}
p=A\rho - \frac{D}{\rho^{\alpha}}
\end{equation}
where $A$ and $D$ are positive constants and $\alpha$ lies within the range $0\leq\alpha\leq 1$. Rewriting Einstein equations (\ref{e1}) and (\ref{e2}), we thus have 
\[
\frac{-1}{2r^2}(f_{ch}+rf'_{ch})=\rho(r)
\]
\begin{equation}
\frac{1}{4r}(2f'_{ch}+rf''_{ch})=\frac{1}{2}(\rho(r)+3p(r))
\end{equation}
in which $f_{ch}$ is used to indicate that we are dealing with a static spherically symmetric black hole surrounded by Chaplygin gas.
These equations together with (\ref{Chaplygin State Equation}) form a closed system of equations for the three unknown functions $f_{ch}$, $\rho$ and $p$. 
After a simple calculation, we find that
\begin{equation}\label{rho_Ch}
\rho = \left[K+\frac{E}{r^l}\right]^m
\end{equation}
\begin{equation}\label{fF}
f_{ch}=-\frac{2r^2}{3K^m} {_2}F_1(\frac{-3}{l}, 
  -m, 1-\frac{3}{l}, -\frac{E}{Kr^l}) - \frac{r_g}{r}
\end{equation}
where
\begin{equation}
K=\frac{3D}{1+A} \hspace{1cm}l=3(1+A)(1+\alpha)\hspace{1cm}m=(1+\alpha)^{-1}\hspace{1cm}
\end{equation}
where $r_g$ is an integration constant and ${_2}F_1$ is hypergeometric function. The first term in (\ref{fF}) arises from the Chaplygin gas and in the second term, a minus sign is chosen to get the standard term of the Schwarzscild metric. For a constant equation of state parameter, $D=0$ , we find that $\rho\sim r^{3(1+A)}$ and the first term of $f_{ch}$ is proportional to $1/r^{(1+3A)}$ which agrees with (\ref{eq: f_q Sum}). Another special case is the pure Chaplygin gas where $A=0$ and $\alpha=1$. For this case, we obtain
\begin{equation}
\rho = \left[D+\frac{E}{r^6}\right]^{1/2}
\end{equation}
and
\begin{equation}
f_{ch}^{(pure)}=-\frac{2r^2}{3}\sqrt{D+\frac{E}{r^6}}+2r^2\frac{\sqrt{E(D+\frac{E}{r^6})}\text{Arctanh}\sqrt{1+\frac{Dr^6}{E}}}{3\sqrt{E+Dr^6}}
\end{equation}
The last interesting case studied here, is the space-time seen by an observer located far away from the black hole surrounded by a general Chaplygin gas. In this case, the second term in \eqref{rho_Ch} is small, one can easily show that   
\begin{equation}
f_{ch}=-2\left(\frac{1+A}{D}\right)^{1+\alpha}\left[\frac{r^2}{3}-\frac{E(1+A)}{D(1+\alpha)}\frac{r^{-3(1+A)(1+\alpha)-1}}{3(\alpha+A+\alpha A)}\right]-\frac{r_g}{r}
\end{equation}
in which the first term on the right hand side  corresponds to the presence of cosmological constant and the third term is the usual Schwarzschild term. Thus for a distant observer who is located far away from the black hole, the presence of the Chaplygin gas is distinguishable from the cosmological constant. 

It should be noted that in derivation of expressions (\ref{rho_Ch}) and (\ref{fF}), we have used a non-perfect Chaplygin gas with anisotropic pressures given by (\ref{pha}). As it is shown in section \ref{Kiselev Space-time}, this is resulted directly from our assumption $g_{tt}g_{rr}=-1$. However, this is not always true. For example, there are some solutions of Tolman-Oppenheimer-Volkov equations in the case of a perfect Chaplygin gas, for which  $g_{tt}g_{rr}\neq-1$ \cite{Gorini:2008zj}. This is also true for some other solutions of Tolman-Oppenheimer-Volkov equations in the presence of a perfect fluid.

\section{Concluding Remarks}\label{Sec: last}
In this paper, we have investigated the deflection angle of light by a quintessential black hole where the quintessence has extremely small deviation from the cosmological constant, $\omega=-1$. This allows us to use a perturbative analysis in which, in addition to the dimensionless Schwarzschild radius, $\bar{r}_g$, a new deviation parameter $\varepsilon$ is appeared. We have obtained the usual expansion of the bending light in Schwarzschild space-time in terms of $\bar{r}_g$ in which any order of expansion is corrected by an expansion in terms of $\varepsilon$. The deflection angle is calculated up to the second order in $\bar{r}_g$ and the first order in $\varepsilon$ and it can be seen that in photon trajectory, the contribution of quintessence can not be absorbed into the definition of impact parameter in contrast to the cosmological constant \cite{Arakida:2011th}. This is also true for light bending angle both in the case of cosmological constant and also quintessence. \\
We also generalized the Kiselev metric to a black hole immersed in a Chaplygin gas. It is worth noting that in Kiselev space-time, a non-perfect energy-momentum tensor acts as the source of Einstein equations. Therefore, this solution is different from the solutions of Tolman-Oppenheimer-Volkov equations when a perfect fluid is included with any equation of state (linear, Chaplygin, etc.) except for the cosmological constant. 
\vglue1cm
{\bf Acknowledgements:}

This work is  supported by a grant from university of Tehran.

\end{document}